\def\Box{{\mathchoice\sqr54\sqr54\sqr33\sqr23}\,}
\def\pr#1 {Phys. Rev. {\bf D#1\tie }}
\def\pe#1 {Phys. Rev. {\bf #1\tie}}
\def\pre#1 {Phys. Rep. {\bf #1\tie}}
\def\pl#1 {Phys. Lett. {\bf #1B\tie }}
\def\prl#1 {Phys. Rev. Lett. {\bf #1\tie }}
\def\np#1 {Nucl. Phys. {\bf B#1\tie }}
\def\ap#1 {Ann. Phys. (NY) {\bf #1\tie }}
\def\cmp#1 {Commun. Math. Phys. {\bf #1\tie }}
\def\imp#1 {Int. Jour. Mod. Phys. {\bf A#1\tie }}
\def\mpl#1 {Mod. Phys. Lett. {\bf A#1\tie}}
\def\zp#1 {Z. Phys. {\bf C#1\tie}}
\def\amp#1 {Adv. Theor. Math. Phys. {\bf#1\tie}}
\def\jhep#1 {JHEP {\bf #1\tie\rm }}
\def\tie{\noexpand~}

\def\s{(\sigma)}

\def\ints{\int d\sigma\,}

\newcommand{\I}{{\rm i}}
\def\ra{\rightarrow}
\def\be{\begin{equation}}
\def\ee{\end{equation}}
\def\bea{\begin{eqnarray}}
\def\eea{\end{eqnarray}}
\def\Ref#1{(\ref{#1})}

\def\nn{\nonumber}

\catcode`@=11
\def\marginnote#1{}
\newcount\hour
\newcount\minute
\newtoks\amorpm
\hour=\time\divide\hour by60
\minute=\time{\multiply\hour by60 \global\advance\minute by-\hour}
\edef\standardtime{{\ifnum\hour<12 \global\amorpm={am}%
        \else\global\amorpm={pm}\advance\hour by-12 \fi
        \ifnum\hour=0 \hour=12 \fi
        \number\hour:\ifnum\minute<10 0\fi\number\minute\the\amorpm}}
\edef\militarytime{\number\hour:\ifnum\minute<10 0\fi\number\minute}
\def\draftlabel#1{{\@bsphack\if@filesw {\let\thepage\relax
   \xdef\@gtempa{\write\@auxout{\string
      \newlabel{#1}{{\@currentlabel}{\thepage}}}}}\@gtempa
   \if@nobreak \ifvmode\nobreak\fi\fi\fi\@esphack}
        \gdef\@eqnlabel{#1}}
\def\@eqnlabel{}
\def\@vacuum{}
\def\draftmarginnote#1{\marginpar{\raggedright\scriptsize\tt#1}}
\def\draft{\oddsidemargin 0.0truein
        \def\@oddfoot{\sl preliminary draft \hfil
        \rm\thepage\hfil\sl\today\quad\militarytime}
        \let\@evenfoot\@oddfoot \overfullrule 3pt
        \let\label=\draftlabel
        \let\marginnote=\draftmarginnote
   \def\@eqnnum{(\theequation)\rlap{\kern\marginparsep\tt\@eqnlabel}%
\global\let\@eqnlabel\@vacuum}  }
\catcode`@=12
\documentstyle[epsf, 12pt]{article}
\begin{document}

\thispagestyle{empty}

\bigskip\bigskip
\begin{center}
\Large{\bf STRINGS IN NONTRIVIAL GRAVITINO\\[2mm]
AND RAMOND-RAMOND BACKGROUNDS}
\end{center}

\vskip 1.0truecm

\centerline{\bf Ioannis  
Giannakis\footnote{giannak@summit.rockefeller.edu}}
\vskip5mm
\centerline{\it Physics Department}
\centerline{\it Rockefeller University}
\centerline{\it 1230 York Avenue}
\centerline{\it New York, NY 10021}

\vskip5mm

\bigskip \nopagebreak \begin{abstract}
\noindent
In this paper we discuss deformations of the BRST operator
of the fermionic string.  These deformations
preserve inlpotency of the BRST operator and correspond to turning
on infinitesimal Gravitino and
Ramond-Ramond spacetime fields.
\end{abstract}

\newpage\setcounter{page}1

\vfill\vfill\break

One of the outstanding problems of string theory is
to understand the equations of motion for the
fields of the theory ( massless and massive ) and the
higher symmetries that relate them \cite{nontrivial},\cite{meov}.
Progress towards this direction can be achieved
by studying infinitesimal deformations of the SuperVirasoro
algebras that preserve superconformal invariance \cite{mg}.
The problem of finding superconformal deformations
is an interesting problem in its own right, but
it also provides us with insights into the
symmetry structure of string theory since spacetime
symmetry transformations are particular superconformal deformations
\cite{mgn}. 
In a recent paper \cite{bg} we constructed a class of superconformal
deformations, termed {\it canonical deformations}, in terms
of superfields (see also \cite{ra}). Although properly speaking
we need to discuss deformations of two
copies of the SuperVirasoro algebra in the remainder of this
paper we shall concentrate only on one copy.
More specifically we found that a deformation
of the form
\bea
{\delta}{\cal T}\s={\delta}T_{F}\s+{\theta}{\delta}T\s
={\Phi}_{F}\s+{\theta}{\Phi}_{B}\s
\label{eqveron}
\eea
where $\Phi_{B}({\Phi}_{F})$ is the bosonic (fermionic)
component of a superfield of dimension $({1\over 2}, {1\over 2})$,
preserves superconformal invariance.

Canonical deformations have a number of interesting features:
superprimary fields of dimension $({1\over 2}, {1\over 2})$ are
in natural correspondence with the physical states of string
theory, being the vertex operators. As such they have a nice
spacetime interpretation in terms of turning on spacetime fields.
Appealing though they are canonical deformations have also
significant drawbacks. They
do not appear to describe spacetime fermions and R-R bosonic
fields which are written in terms of spin fields. Spin fields cannot be
written as superfields. These string backgrounds have
attracted interest recently due to the conjectured AdS/CFT
equivalence \cite{mal}.
We might attempt to identify
the bosonic component $\Phi_{B}(\sigma)$ of the
canonical deformation with the appropriate spacetime gravitino
vertex operator
\bea
\delta T(\sigma)&=&\Phi_{B}(\sigma)=
{\Psi}_{\mu}^{\alpha}(X)S_{\alpha}
e^{-{{\phi}\over 2}}{\overline{\partial}}X^{\mu}
+{\tilde\Psi}_{\mu}^{\alpha}(X)(X){\tilde S}_{\alpha}
e^{-{{\tilde{\phi}}\over 2}}{\partial}X^{\mu} \nn\\[0mm]
&+&{\partial_{\lambda}}{\Psi}_{\mu}^{\alpha}(X)S_{\alpha}
e^{-{{\phi}\over 2}}{\tilde{\psi}}^\lambda{\tilde{\psi}}^\mu
+{\partial_{\lambda}}{\tilde{\Psi}}_{\mu}^{\alpha}(X)
{\psi}^\lambda{\psi}^\mu{\tilde S}_{\alpha}
e^{-{{\tilde{\phi}}\over 2}}. \nn\\
\label{eqaboua}
\eea
In order to calculate ${\delta}T_F$ we need to calculate the commutator
of $\Phi_{B}(\sigma)$ with the supercurrent $T_F(\sigma)$.
The commutator of the vertex operator which is
written in terms of spin fields with
the supercurrent $T_F$ is not well-defined since the corresponding
OPE in the complex plane involves branch cut singularities
\bea
T_{F}(z){\Phi_B}(w)&=&{{\gamma^{\lambda}}_{\alpha{\dot{\beta}}}
{\partial_{\lambda}}{\Psi}_{\mu}^{\alpha}(X)S^{\dot\beta}
e^{-{{\phi}\over 2}}{\overline{\partial}}X^{\mu} \over {(z-w)^{3\over 2}}}
+{{\gamma^{\lambda}}_{\alpha{\dot{\beta}}}
{\Psi}_{\mu}^{\alpha}(X)S^{\dot\beta}
e^{-{{\phi}\over 2}}{\partial}X_{\lambda}
{\overline{\partial}}X^{\mu} \over {(z-w)^{1\over 2}}} \nn\\[0mm]
&+&{{\gamma^{\rho}}_{\alpha{\dot{\beta}}}
{\partial_\rho}{\partial_{\lambda}}
{\Psi}_{\mu}^{\alpha}(X)S^{\dot\beta}
e^{-{{\phi}\over 2}}
{\tilde{\psi}}^\lambda{\tilde{\psi}}^\mu \over {(z-w)^{3\over 2}}}
+{{\gamma^{\rho}}_{\alpha{\dot{\beta}}}
{\partial_\lambda}{\Psi}_{\mu}^{\alpha}(X)S^{\dot\beta}
e^{-{{\phi}\over 2}}{\partial}X_{\lambda}
{\tilde{\psi}}^\lambda{\tilde{\psi}}^\mu \over {(z-w)^{1\over 2}}} \nn\\
\label{eqtzenam}
\eea
where we have omitted terms that are either regular or
have poles as singularities. 
This suggests then that the canonical
deformations we have constructed in terms of superfields
are not the most general solution to the deformation equations.

The failure to derive an expression for $\delta T_F$
is puzzling.  It is not clear if the presence of the
spin fields breaks superconformal invariance or if
superconformal invariance is realised in an apparent nonlocal manner.

Instead of deforming the stress energy superfield,
we could have deformed the BRST charges $Q$ and
${\overline Q}$.  Nilpotency then requires
\be
\lbrace Q, \delta Q \rbrace \,=\,0\ , \quad \lbrace
{\overline Q},
\delta {\overline Q} \rbrace\,=\,0\ , \quad \lbrace Q,
\delta {\overline Q}
\rbrace\,+\,\lbrace {\overline Q}, \delta Q \rbrace
\,=\,0
\label{eqkwmh}
\ee
under the infinitesimal deformations
\be
Q \,\rightarrow\, Q\,+\,\delta Q\ , \qquad {\overline Q}
\,\rightarrow\,{\overline Q}\,+\,\delta {\overline Q}\ .
\label{eqkrokos}
\ee

Although the two approaches are equivalent in the presence
of NS-NS backgrounds, they are not necessarily equivalent
in the presence of gravitino and Ramond-Ramond backgrounds.
In fact, given
a deformed BRST charge, the components of the deformed
stress energy superfield can be extracted by calculating
the commutator or anticommutator of $Q$ with the ghost field
$b$ or $\beta$, assuming that
the commutator or anticommutator
exists. In the presence of spin fields
the commutator of $\beta$ with
the deformed BRST charge does not exist.

Next we shall derive the form of the deformation of the
BRST operator which corresponds to turning on
a spacetime gravitino. We shall employ
a particular formalism which relates
superconformal deformations and spacetime symmetries.
In string theory, the stress energy superfield
${\cal T}_\Phi
= T_{F(\Phi)} + \theta T_{\Phi}$ depends on the spacetime
fields $\Phi$.  Spacetime symmetries are superconformal
deformations that induce changes in the spacetime fields:
\bea
\delta T &=& \I\, [h, T_{\Phi}]
\ =\ T_{\Phi+{\delta{\Phi}}}\,-\,T_{\Phi}  \nn\\[0mm]
\delta T_F &=& \I\, [h, T_{F(\Phi)}]
\ =\ T_{F(\Phi+{\delta{\Phi}})}
\,-\,  T_{F(\Phi)} \ .
\label{eqlouc}
\eea
The operator $h$ is the generator of the spacetime symmetry;
it is the zero mode of a sum of dimension $(1,0)$ and $(0,1)$
currents.
The previous discussion can also be carried through in terms
of the BRST formalism.  Let us suppose that $Q_\Phi$
is nilpotent BRST charge, function of
the spacetime fields. Then $\Phi
\ra {\Phi+{\delta{\Phi}}}$ is a spacetime symmetry if
\bea
\delta Q_\Phi = \I\, [ h, Q_\Phi ] \ =\
Q_{\Phi+{\delta{\Phi}}}\,-\, Q_\Phi
\label{eqlucio}
\eea

In order to
generate a gravitino background
we need to perform a supersymmetry transformation
about flat spacetime. The operator $h$ that generates 
$N=2$ spacetime
supersymmetry transformations is \cite{susy}
\bea
h=\ints \Big [{\epsilon^{\alpha}}(X)S_{\alpha}e^{-{{\phi}\over 2}}
+{\tilde{\epsilon}}^{\alpha}(X)
{\tilde S}_{\alpha}e^{-{{\tilde{\phi}}\over 2}} \Big ]
\label{eqafddcm}
\eea
where $S^{\alpha}$, $e^{-{{\phi}\over 2}}$,
${\tilde S}^{\alpha}$ and $e^{-{{\tilde{\phi}}\over 2}}$
are the spin fields
for the two-dimensional fermions $\psi_\mu(\sigma)$,
$\tilde\psi_\mu(\sigma)$ and
the superconformal ghosts $\beta(\sigma), \gamma(\sigma)$
$\tilde\beta(\sigma), \tilde\gamma(\sigma)$
respectively.
The integrand is again superprimary of dimension $(1, 0)$ only if the
parameters $\epsilon^{\alpha}(X)$ and $\tilde\epsilon^{\alpha}(X)$
satisfy
\bea
\Box\epsilon^{\alpha}(X)=\Box\tilde\epsilon^{\alpha}(X)=0,
\qquad {\gamma}^{\mu}{\partial_{\mu}}
{\epsilon^{\alpha}}={\gamma}^{\mu}{\partial_{\mu}}
{\tilde\epsilon}^{\alpha}=0.
\label{eqaseme}
\eea
Let's calculate $i[h, Q]$. The result is
\bea
\delta Q=i[h, Q]&=&\ints \Big (c{\partial_{\mu}}
{\epsilon^{\alpha}}(X)S_{\alpha}
e^{-{{\phi}\over 2}}{\overline{\partial}}X^{\mu}
+c{\partial_{\mu}}
{\tilde\epsilon}^{\alpha}(X){\tilde S}_{\alpha}
e^{-{{\tilde{\phi}}\over 2}}{\partial}X^{\mu}\Big )\s \nn\\[0mm]
&+&{1\over 2}\ints e^{\phi}{\eta}{\partial_{\mu}}
{\tilde\epsilon}^{\alpha}(X){\psi^\mu}{\tilde S}_{\alpha}
e^{-{{\tilde{\phi}}\over 2}} \s. \nn\\
\label{eqwascd}
\eea
Although these
backgrounds are pure gauges, we gain an insight
into the form of the deformation which corresponds to turning
on the appropriate spacetime fields.
The obvious ansatz for the canonical deformation then corresponding to
gravitino propagation about flat spacetime is
\bea
\delta Q=\ints \Big [ c
{\Psi}_{\mu}^{\alpha}(X)S_{\alpha}
e^{-{{\phi}\over 2}}{\overline{\partial}}X^{\mu}
+c{\tilde\Psi}_{\mu}^{\alpha}(X)
(X){\tilde S}_{\alpha}
e^{-{{\tilde{\phi}}\over 2}}{\partial}X^{\mu}
+{1\over 2}e^{\phi}{\eta}{\tilde\Psi}_{\mu}^{\alpha}(X)
{\psi^\mu}{\tilde S}_{\alpha}
e^{-{{\tilde{\phi}}\over 2}} \Big ]\s
\label{eqwabcd}
\eea
since under a supersymmetry transformation the gravitinos transform
about flat spacetime as,
$\delta {\Psi}_{\mu}^{\alpha}={\partial_{\mu}}{\epsilon^{\alpha}}$
and $\delta {\tilde{\Psi}}_{\mu}^{\alpha}=
{\partial_{\mu}}{\tilde\epsilon}^{\alpha}$.
This particular ansatz does not obey the deformation equations
\bea
\{Q, \delta Q\}=0, \qquad \{ {\overline Q}, {\delta}{\overline Q}\}=0,
\qquad \{Q, \delta {\overline Q}\}+\{\delta Q, {\overline Q}\}=0
\label{eqzouni}
\eea
and it has to be supplemented with extra terms. We find that
\bea
\delta Q&=&\ints \Big [ c
{\Psi}_{\mu}^{\alpha}(X)S_{\alpha}
e^{-{{\phi}\over 2}}{\overline{\partial}}X^{\mu}
+c{\tilde\Psi}_{\mu}^{\alpha}(X)
(X){\tilde S}_{\alpha}
e^{-{{\tilde{\phi}}\over 2}}{\partial}X^{\mu}
+c{\partial_{\lambda}}{\Psi}_{\mu}^{\alpha}(X)S_{\alpha}
e^{-{{\phi}\over 2}}{\tilde{\psi}}^\lambda{\tilde{\psi}}^\mu \nn\\[0mm]
&+&c{\partial_{\lambda}}{\tilde{\Psi}}_{\mu}^{\alpha}(X)
{\psi}^\lambda{\psi}^\mu{\tilde S}_{\alpha}
e^{-{{\tilde{\phi}}\over 2}}+
{1\over 2}e^{\phi}{\eta}{\tilde\Psi}_{\mu}^{\alpha}(X)
{\psi^\mu}{\tilde S}_{\alpha}
e^{-{{\tilde{\phi}}\over 2}}+ \Big ]\s
\label{eqzeff}
\eea
satisfy equations \Ref{eqzouni} if the
gravitino wave function satisfies the following equations
\bea
\Box  {\Psi}_{\mu}^{\alpha} (X)=0, \quad {\gamma}^{\mu}{\partial_{\mu}}
{\Psi}_{\nu}^{\alpha} (X)=0, \quad {\partial^{\mu}}
{\Psi}_{\mu}^{\alpha} (X)=0. 
\label{eqasxiuf}
\eea
We observe the emergence of an equation of motion and
a gauge condition. Of course the first equation is redundant
since it follows from the Dirac equation. The gauge condition
imposes transversality in $X$-space and eliminates
part of the spin $1\over 2$ component of ${\Psi}_{\mu}^{\alpha}(X)$.
The gauge condition does not fix the gauge completely and
subsequently the fermionic wavefunction describes the
emission of a gravitino (spin $3\over 2$ part) and a dilatino
(spin $1\over 2$ part). In order to separate the gravitino
and dilatino parts we write ${\Psi}_{\mu}^{\alpha}(X)=
{\chi}_{\mu}^{\alpha}(X)+{\gamma_{\mu}}{\lambda}^{\alpha}(X)$
and demand that $\gamma^{\mu}{\chi}_{\mu}^{\alpha}=0$. The
gravitino and dilatino wavefunctions can be expressed in
terms of  ${\Psi}_{\mu}^{\alpha}(X)$
\bea
{\lambda}^{\alpha}(X)={1\over D}{\gamma^{\mu}}{\Psi}_{\mu}^{\alpha}(X),
\qquad {\chi}_{\mu}^{\alpha}(X)={\Psi}_{\mu}^{\alpha}(X)-
{1\over D}{\gamma_{\mu}}{\gamma^{\lambda}}{\Psi}_{\lambda}^{\alpha}(X)
\label{eqcbxpo}
\eea
and the equations \Ref{eqasxiuf} imply
\bea
{\gamma}^{\mu}{\partial_{\mu}}
{\chi}_{\nu}^{\alpha} (X)=2{\partial_{\nu}}{\lambda^\alpha},
\quad \gamma^{\mu}{\chi}_{\mu}^{\alpha}(X)=0,
\quad {\partial^{\mu}}
{\chi}_{\mu}^{\alpha} (X)=0, \quad {\gamma}^{\mu}{\partial_{\mu}}
{\lambda}^{\alpha} (X)=0.
\label{eqasxiuf}
\eea
It is obvious again that the most general superconformal deformation
is not canonical since it corresponds to turning on spacetime
fields in a particular gauge. In a subsequent publication
we shall discuss
how to relax the gauge condition and go beyond canonical deformations.

Finally we will discuss deformations that correspond to turning on
Ramond-Ramond spacetime fields. These fields
appear in type $II$ superstrings
and their vertex operators are written in
terms of bispinors $F^{\alpha \beta}(X)$.
The spinor indices are contracted by the left-right spin fields
\cite{dol}.
Strings in Ramond-Ramond backgrounds have been discussed
in \cite{ram}.
In order to generate
a Ramond-Ramond background about flat spacetime we need to perform
two consecutive supersymmetry transformations. This leads us
\Ref{eqwabcd} to the following ansatz for the superconformal deformation
that corresponds to turning on Ramond-Ramond backgrounds
\bea
\delta Q=\ints c(\sigma)
F^{\alpha\beta}(X)S_{\alpha}{\tilde S}_{\beta}
e^{-{1\over 2}({\phi}+{\tilde{\phi}})}.
\label{eqwabefl}
\eea
This ansatz obeys the deformation equations
if the bispinor $F^{\alpha\beta}(X)$ obeys the following equations
\bea
\Box  F^{\alpha\beta} (X)=0, \quad
{\gamma}^{\mu}_{\alpha{\dot{\beta}}}{\partial_{\mu}}
F^{\alpha\beta} (X)=0 \quad
{\gamma}^{\mu}_{\beta{\dot{\beta}}}{\partial_{\mu}}
F^{\alpha\beta} (X)=0.
\label{eqsvoiuf}
\eea
The bispinors $F^{\alpha\beta}$ describe a collection of massless
antisymmetric tensors $F^{\mu_{1}\mu_{2} \cdots \mu_{d}}$ as can
be seen by expanding them in a complete basis of all gamma matrix
antisymmetric products
\bea
F^{\alpha\beta} (X)={\sum_{n=0}^{10}}{{i^n}\over {n!}}
F^{\mu_{1}\mu_{2} \cdots \mu_{n}}(X)
({\gamma}_{\mu_{1}\mu_{2} \cdots \mu_{n}})^{\alpha\beta}.
\label{amazon}
\eea
The chirality
conditions that the bispinor obeys limits the number of the antisymmetric
tensors present in the spectrum of the theory.
In type $IIA$ string theory $S_{\alpha}e^{-{{\phi}\over 2}}$
and ${\tilde S}_{\beta}e^{-{{\tilde\phi}\over 2}}$ have opposite
chirality while in type $IIB$ the have the same
\bea
(\gamma_{11})^{\alpha}_{\delta}F^{\delta\beta}(X)={\pm}
F^{\alpha\delta}(X)(\gamma_{11})^{\alpha}_{\delta}=F^{\alpha\beta}(X).
\label{barnes}
\eea
Furthermore we can
convert the equations of motion for the bispinor wavefuction onto equations
for the antisymmetric tensor wavefunctions by using $\gamma$ matrix
identities
\bea
{\partial^{[{\lambda}}}F^{\mu_{1} \cdots \mu_{d}]}=0 \qquad
{\partial_{\lambda}}F^{\lambda\mu_{2} \cdots \mu_{d}}=0
\label{eqxojhf}
\eea
which are the Bianchi identity and the massless equation of motion
for an antisymmetric tensor field strength.

In this paragraph we shall summarize what we have done in this
paper. We discussed deformations of superconformal field theories
by varying the BRST operator $Q$ such that
$(Q+\delta Q)^2=0$, thus preserving nilpotency
of the BRST operator to first order in $\delta Q$. These
deformations describe strings propagating in
gravitino and Ramond-Ramond backgrounds.

\section{Acknowledgments}

This paper is based on work I have done in collaboration
with J. Bagger.
This work was supported in part by the Department of Energy Contract
Number DE-FG02-91ER40651-TASKB.

\end{document}